# Nonequilibrium dynamics of membraneless active droplets

Chenxi Liu[1], Ding Cao[1], Siyu Liu [1], Yilin Wu[1*]

[1] *The Chinese University of Hong Kong, Shatin, NT, Hong Kong, P.R. China.*

*To whom correspondence should be addressed.  Mailing address: Room 306, Science Centre, Department of Physics, The Chinese University of Hong Kong, Shatin, NT, Hong Kong, P.R. China.  Tel: (852) 39436354.  Fax: (852) 26035204. Email: ylwu@cuhk.edu.hk

**Abstract**

Membraneless droplets or liquid condensates formed via liquid-liquid phase separation (LLPS) play a pivotal role in cell biology and hold potential for biomedical engineering.  While membraneless droplets are often studied in the context of interactions between passive components, it is increasingly recognized that active matter inclusions, such as molecular motors and catalytic enzymes in cells, play important roles in the formation, transport and interaction of membraneless droplets. Here we developed a bacteria-polymer active phase separation system to study the nonequilibrium effect of active matter inclusions on the LLPS dynamics.  We found that the presence of bacterial active matter accelerated the initial condensation of phase-separated liquid droplets but subsequently arrested the droplet coarsening process, resulting in a stable suspension of membraneless active droplets packed with motile bacterial cells.  The arrested phase separation of the bacterial active droplet system presumably arises from anti-phase entrainment of interface fluctuations between neighboring droplets, which reduces the frequency of inter-droplet contact and suppresses droplet coarsening.  In addition, the active stresses generated by cells within the droplets give rise to an array of nonequilibrium phenomena, such as dominant long-wavelength fluctuations and enhanced droplet transport with short-term persistent motion due to spontaneous symmetry breaking. Our study reveals a unique mechanism for arrested phase separation and long-term stability in membraneless droplet systems.  The bacteria-polymer active phase separation system opens a new avenue for studying the dynamics of membraneless active droplets relevant to non-equilibrium LLPS in cells and in biomedical engineering applications.

**Introduction**

Liquid-liquid phase separation (LLPS) is a process where a mixture of different components spontaneously segregates into distinct liquid phases (1, 2). Membraneless droplets or liquid condensates formed via LLPS play a pivotal role in cell biology (3, 4) and hold potential for biomedical engineering (5, 6). In living cells, such liquid condensates modulate cellular organization and functions by providing a locally distinct molecular environment while allowing molecular exchange via diffusion. In biomedical engineering, synthetic biopolymer condensates can be used for biocompatible therapeutic delivery (7, 8), biomaterial self-assembly (9-11), and tissue engineering (12). Moreover, liquid condensates have been suggested to resemble prebiotic protocells prior to the emergence of the first cell-like entities (13), thus understanding the nonequilibrium dynamics of liquid condensates may shed light into the emergence of life (14, 15).

While liquid-liquid phase separation was often studied in the context of interactions between passive components (1, 2) it is increasingly recognized that active matter inclusions and nonequilibrium processes play important roles (4, 14-17). Living cells are intrinsically out-of-equilibrium and their cytoplasm is replete with active matter units such as molecular motors (18) and catalytic enzymes (19, 20). These active matter inclusions in the cytoplasm of cells influence the dynamics of phase-separated membraneless organelles that perform essential living functions. For example, in the cytoplasm of animal cells, activities of molecular motors and ATP-driven enzymes modulate the dynamics of nucleoli (21) and stress granules (22, 23), both of which are RNA-protein liquid condensates; in bacterial cells, the molecular motor ParA prevents the fusion of liquid condensates that are involved in DNA segregation and compartmentalization (24). In LLPS systems infused with reconstituted cytoskeletal molecular motors, the activity of self-propelled constituents was recently shown to alter the phase diagram and interfacial dynamics of passive condensates (9, 25-27). However, it is still largely unclear how active stresses generated *from within* by active matter inclusions impact liquid condensate formation, transport and interactions. In particular, many types of membraneless organelles in living cells can resist the coarsening process driven by free energy minimization (28-30), and it is an open question how active matter inclusions may contribute to the stability of membraneless organelles.

Here we introduce an active LLPS system consisting of motile bacteria—a premier model system of active matter—dispersed in an aqueous solution of two well-characterized phase-separating polymers, i.e., gelatin and dextran. We discovered that the presence of motile bacteria accelerated the initial condensation of membraneless droplets but subsequently arrested the droplet coarsening process, resulting in a stable suspension of membraneless active droplets packed with motile bacterial cells. The active stresses generated by cells within the droplets give rise to an array of nonequilibrium phenomena, such as dominant long-wavelength fluctuations and enhanced droplet transport with short-term persistent motion due to spontaneous symmetry breaking.  Moreover, the active droplets displayed anti-phase entrainment of interface fluctuations that suppressed droplet coarsening, offering a unique mechanism to arrest phase separation and to promote long-term stability of membraneless droplet systems.  The bacteria-polymer active phase separation system presented here provides a platform for studying the dynamics of membraneless active droplets.  As a scaled-up analogue of enzyme-biopolymer composites ubiquitous in living cells, it may help to understand non-equilibrium LLPS in cells and to engineer synthetic membraneless droplet systems for biomedical applications.

**Results**

**Motile bacteria accelerate initial droplet condensation but suppress droplet coarsening**

To study the effect of active matter inclusion on the dynamics of liquid-liquid phase separation, we mixed motile cells of the model flagellated bacterium *Bacillus subtilis* (31, 32) into the dextran-gelatin LLPS system (with gelatin being the low-volume-fraction component).  The mixture was initially homogenized by vortexing and then injected into a closed fluid chamber (Methods).  In the absence of motile bacteria, the dextran-gelatin mixtures undergo phase separation driven by entropic forces upon reaching critical concentrations (33), forming gelatin-rich droplets dispersed in the dextran phase (Fig. S1; Methods) as visualized by the fluorescent label Oregon Green™ 488 gelatin conjugate (Fig. S2, left; Methods). Upon mixing, *B. subtilis* cells were enriched in the gelatin-rich phase by ~11-fold of that in the dextran phase (Methods), forming membraneless active droplets packed with motile bacteria (with a droplet diameter of tens of μm and a cell density up to the order of ~$10^{10}$ cells/mL) (Fig. 1A; Fig. S2, right; Movie 1, Movie 2).  Cells exhibited vigorous motility within both gelatin-rich and dextran-rich phases, and they can freely cross the phase boundary (Movie 3, Movie 4).  The enrichment of cells in the gelatin-rich phase, as manifested by the higher cell fluxes entering versus leaving the gelatin-rich phase (Fig. 1B; Methods), is presumably due to electrostatic interactions between negatively charged cells and the positively charged gelatin polymers under our pH=7.0 condition (34); chemical affinity between cells and polymers was also reported to affect bacterial partitioning in another aqueous liquid-liquid phase separation system (dextran and polyethylene glycol) (35).  Indeed, passive microspheres with negative charge were also spontaneously enriched in the active droplets (Fig. 1C, D); this fact allowed us to label the active droplets with photostable fluorescence microspheres and to follow their dynamics under confocal microscopy over a long period of time without photobleaching (Fig. S3; Methods).

Surprisingly, we found that the membraneless active droplets encapsulating motile bacteria had a size up to one order of magnitude greater than that of the passive counterparts without motile bacteria (Fig. 2A, red data points versus the black data points).   While their size was positively correlated with average cell density in the suspension (Fig. 2A), the bacterial active droplets typically had a diameter of tens of µm.   Interestingly, the initial condensation process of bacterial active droplets occurred so rapidly that it had been nearly completed within ~1 min during sample loading prior to microscopy observation (Methods); throughout the entire imaging period, the size of active droplets only grew slightly for the first ~20 min and then it remained constant (Fig. 2B, upper; Fig. 2C), signifying arrested phase separation at the stage of droplet coarsening.   By contrast, the size of passive gelatin-rich droplets formed in the absence of bacterial inclusion was up to several µm only, and it increased over time with a $\sim t^{1/3}$ scaling consistent with diffusion-driven droplet coarsening (36, 37) (Fig. 2B, lower).   We note that cell motility is essential for the dramatic size difference between active and passive gelatin-rich droplets, since non-motile *B. subtilis* cells mixed into the gelatin-dextran solution (Methods) resulted in droplets of a size similar to that of the bacteria-free droplets (Fig. 2A, blue data points).

Taken together, the results presented above showed that motile bacteria accelerated the initial condensation but subsequently arrested the droplet coarsening.   The underlying mechanism for the arrested droplet coarsening is intriguing, as it may offer a means to stabilize suspensions of membraneless droplets in biomedical applications (7, 8) and may help to explain the stability of membraneless organelles in living cells (28-30).   Following the line of investigation in our study, we would first describe the behavior of individual bacterial active droplets and return to the mechanism of arrested phase separation later.   First of all, we sought to understand how the presence of motile bacteria accelerated the initial condensation of the active droplets.   By examining the spatial distribution of gelatin-rich phase during condensation process (Methods), we found that most isolated cells in the suspension were associated with a blob of gelatin polymers (Fig. 2D).   This phenomenon presumably resulted from the preferential adhesion of gelatin-rich droplets onto cell surface that has higher affinity with gelatin molecules due to electrostatic interactions; that is, cells were carrying a gelatin blob as cargo while swimming around.   When a cell entered a gelatin-rich active droplet packed with other cells, the gelatin blob it carried became part of the active droplet.   Due to the unbalanced cellular fluxes entering versus exiting active droplets in the initial growth stage (Fig. 1B), the volume of active droplets would increase at a much faster rate than in the passive case without motile cells.   The result suggests that bacterial motility accelerates initial liquid droplet condensation via active transport of the higher-affinity polymer component.   This mechanism of activity-driven condensation may be relevant to liquid condensation processes in living cells where enzyme-biopolymer complexes are ubiquitous.

**Nonequilibrium interface fluctuation of bacterial active droplets exhibits a generic scaling behavior**

The bacteria-laden membraneless active droplets displayed an array of nonequilibrium phenomena not seen in the passive counterpart. A prominent observation is that the active droplets displayed substantial interface fluctuations due to active stresses generated by cell motility (Fig. 3A; Movie 5 and 6). Following a previous study on giant unilamellar vesicles (GUVs) encapsulating synthetic self-propelled particles (38), we decomposed the interface fluctuations at an instant into fluctuation modes by analyzing the two-dimensional contour of the active droplets (Methods), with the $l$-th mode amplitude $\alpha_l$ given by the following expression:

$$\alpha_l = \frac{1}{n} \sum_{m=0}^{n-1} r(\theta_m) \exp\left(-\frac{2\pi i l m}{n}\right)$$

, where $l$ is the mode number, $r(\theta_m)$ is the contour radial position at the $m$-th angular position $\theta_m = \frac{2\pi m}{n}$ in polar coordinates, $n$ is the number of chosen angular positions, and $m$ is the index of $n$.

For passive gelatin-rich droplets without bacteria, the interface fluctuation spectra $\langle \alpha_l^2 \rangle$ scales approximately as $l^{-1}$ and $l^{-3}$ for small ($l < 20$) and moderate mode numbers (Fig. S4), respectively, which is similar to the cross-sectional contour fluctuations observed in passive membrane vesicles (38, 39). By contrast, the interface fluctuation spectra of membraneless active droplets containing motile bacteria scaled with $l$ approximately as $l^{-3}$ in both the low and moderate mode number regimes (Fig. 3B, data represented by triangles). The ~$l^{-3}$ power law scaling with mode number remained approximately the same when we varied the overall strength of active stresses (by tuning cell number density inside the active droplets) or increased the localization of active stresses (by increasing cell length to induce local polar ordering; Methods), despite that the fluctuation magnitude was higher in both cases (Fig. 3B; Fig. 3A, lower). These results showed that the membraneless active droplets display dominant low-wavenumber (or long-wavelength) interface fluctuations. The dominance of low-wavenumber interface fluctuations as well as the ~$l^{-3}$ scaling in our membraneless active droplets is again similar to that observed in membrane-bound vesicles encapsulating active particles (38-40). Despite the fact that a lipid membrane and the aqueous gelatin-dextran interface both possess ultralow interfacial tensions (on the order of 10 µN/m or less) and finite bending rigidities (41), there are essential differences in terms of compositions, permeability, and stretchability. The similarity in fluctuation spectra highlights the generic nature of nonequilibrium interface fluctuations generated by active stresses.

**Active matter inclusions enhance droplet transport**

Next we examined the spontaneous motion and transport of the membraneless active droplets, a property that has been predicted by various theories (42-44). We measured the mean square displacement (MSD) of our active droplets of different sizes and confirmed a universal short-term persistent motion (lasting for several seconds) followed by a long-term diffusive motion (>~10 s) (Fig. 3C). We noticed that the time scale of membraneless active droplets' persistent motion (several seconds; Fig. 3C) was comparable to the characteristic time scale of the non-equilibrium shape deformations (~6.5 s; Fig. S5).

To understand the connection between the short-term persistent motion of active droplets and the shape deformations, we defined a deformation index $D = 1 - 4\pi S/P^2$ as a proxy for the extent of shape deformations, where $S$ is the area of active droplets in the imaging plane and $P$ is the perimeter, with $C \equiv 4\pi S/P^2$ being the circularity of the active droplet (Methods). Cross-correlation function between the instantaneous displacement and the deformation index of active droplets peaked at zero lag time (Fig. 3D; Methods), showing that the persistent motion coincided with the shape deformation. This coincidence suggests that the short-term persistent motion and shape deformation share the same root in the active stress state of the membraneless active droplet. It's noteworthy that the short-term persistent motion of our membraneless active droplets occurs in bulk liquids as a result of spontaneous symmetry breaking; it does not require contact with a solid surface, which is in contrast to the persistent motion found in several other types of active droplets enclosing active nematics, motile bacteria, or colloidal Quincke rollers (45-47). Interestingly, persistent motion arising from spontaneous symmetry breaking in bulk liquids was also observed in GUVs encapsulating motile bacteria (48); in this peculiar case, activity of motile cells deformed the GUVs drastically, producing long tubular protrusions that served as helical propellers.

In the diffusive regime, the membraneless active droplets had effective diffusivities (denoted as $D_{\text{eff}}$) three orders of magnitude higher than that expected for passive droplets of the same size following Stokes–Einstein relation (Fig. 3E). In contrast to passive objects whose diffusion is driven by thermal fluctuations, the diffusion of active droplets here is driven by fluctuations of bacterial activity that give rise to the short-term persistent motion. Our data showed that the effective diffusivity $D_{\text{eff}}$ can be well fitted to $D_{\text{eff}} \sim R^{-1/4}$ (Fig. 3E). To understand this empirical relation, we note that the characteristic energy associated with active fluctuations is $\Delta E \sim E^{1/2}$, where $E$ is the total active energy input that scales with the number of cells in a droplet $N$ as $E \sim N^{1/2}$(49), so we have $\Delta E \sim N^{1/4}$ or $\Delta E \sim R^{3/4}$ where $R$ is the radius of the active droplet. The relation $D_{\text{eff}} \sim R^{-1/4}$ implies that $D_{\text{eff}} \sim \Delta E/(6\pi \eta R)$, which suggests the presence of a fluctuation-dissipation type relation between the fluctuation energy $\Delta E$ and the effective diffusivity $D_{\text{eff}}$. Indeed, it was showed that fluctuation-dissipation type relations could still hold under certain specific conditions in active matter systems

(50).

**Anti-phase entrainment of interface fluctuations stabilizes the active droplet system**

Intuitively, the enhanced transport of active droplets would expedite droplet coarsening by enhancing the frequency of droplet contact and coalescence. However, according to Fig. 2A, the coarsening of bacterial active droplets appeared to be suppressed by the activity of cells.   To understand these seemingly contradicting results, we studied the interaction between membraneless active droplets. Surprisingly, for two neighboring active droplets, we found that their interface fluctuation patterns were entrained near the gap region in an anti-phase manner when they were at close proximity (Fig. 4A, upper; Movie 6).   For instance, as shown in the image sequence of the upper row of Fig. 4A, the boundary moving velocities of the two droplets near the gap region were strongly correlated, with one advancing and the other retracting or vice versa, reminiscent of the movement in Tango dancing.   The anti-phase entrainment of interface fluctuations between nearby active droplets reduced the frequency of inter-droplet contacts, which presumably suppressed droplet coarsening as shown in Fig. 2B.   This notion is supported by a stochastic simulation of droplet coarsening dynamics (Fig. 4B; Methods).

We note that the long-wavelength fluctuation of membraneless active droplets appeared to coincide with collective cellular motion normal to the droplet boundary (Fig. 4A, lower), suggesting inter-droplet correlation of collective cellular motion.   To examine this idea, we calculated the inter-droplet correlation of bacterial collective velocity field defined as $\rho_{v,v'}(d) = \langle\frac{v(r)\cdot v'(r')}{|v(r)||v'(r')|}\rangle$, where $v(r)$, $v'(r')$ are bacterial collective velocity fields measured within each of two neighboring active droplets by particle image velocimetry, and the angular bracket represents averaging over all positions within each droplet satisfying $|r - r'| = d$ and over time (Methods). Indeed, the bacterial collective velocity field showed strong correlation between two neighboring droplets (Fig. 4C), with a correlation length of 9.5±2.5 (mean±s.d.; N=3) μm for relatively large active droplets (radius ~35 μm).   In addition, the centroid motion of an arbitrary pair of active droplets in the system displayed higher correlation at smaller inter-droplet front separation (i.e., the width of the gap region between two neighboring droplets) (Fig. 4D; Methods).   These results showed that the collective cellular motion of nearby active droplets was mutually entrained or coordinated due to hydrodynamic interactions across their gap region.   We conclude that the coordinated cellular motion between nearby droplets underlies the anti-phase entrainment of their interface fluctuations.

Although the anti-phase entrainment of interface fluctuations between nearby active droplets reduced the frequency of inter-droplet contacts, droplet contact and coalescence could be observed occasionally (Fig. 4E).   These rare events allowed us to investigate how activity affects the temporal dynamics of droplet coalescence in the presence of nonequilibrium activities (Movie 7).  We found that the growth of droplet neck width (denoted as $w$; Methods) during active droplet coalescence exhibited a universal scaling behavior $w \sim t^{1.0}$ independent of droplet size (Fig. 4F). This scaling is faster than that of neck growth during the coalescence of passive gelatin-rich droplets, which followed $w \sim t^{0.5}$ (Fig. S6) (51).  This result suggests that, upon successful droplet contact, active matter inclusions accelerate droplet coalescence, presumably due to activity-induced directed flows that facilitate mass transfer (52).

**Discussion**

In summary, we have developed a simple, generic active phase separation system consisting of motile bacteria dispersed in an aqueous two-phase polymer solution. The motile bacterial cells accelerated the initial condensation of the higher-affinity polymer droplets via an active cargo-transport mechanism, and subsequently stabilized the active phase separation system by suppressing further droplet coarsening.  The arrested droplet coarsening presumably arises from anti-phase entrainment of interface fluctuations between neighboring droplets, which suppresses droplet coarsening by reducing the frequency of inter-droplet contact.  This mechanism is different from the recently reported routes of activity-suppressed phase separation in active-passive mixtures, where interfacial active stresses cause droplet splitting (26) and activity-induced stirring shifts the critical point (53).  Moreover, the active stresses generated by cells within the bacterial active droplets give rise to an array of intriguing nonequilibrium dynamics, such as dominant long-wavelength fluctuations with a generic scaling behavior and enhanced droplet transport with short-term persistent motion in bulk liquids; the latter arose from spontaneous symmetry breaking and was coupled to the shape deformation of active droplets.

Our study reveals a unique mechanism for arrested phase separation and long-term stability in membraneless active droplet systems, which is relevant to non-equilibrium LLPS in cells and in biomedical engineering applications.  It is known that membraneless organelles in living cells can resist coarsening (28), but the underlying mechanisms for the stability are complex and may depend on various factors, such as nonequilibrium chemical reactions (54), reduced surface tension due to protein absorption (29), or passive effect of the local mechanical environment (30).  The mechanism we report here, i.e., anti-phase entrainment of interface fluctuations, may contribute to the stability of membraneless organelles infused with molecular motors and catalytic enzymes.  Meanwhile, in engineering applications, it is often desirable to have a stable suspension of membraneless droplets; for instance, optimization of droplet size and stability is a critical factor for efficient therapeutic delivery (7, 8).  Our work suggests that active matter inclusions could become useful tools for the regulation of size distribution and stability of synthetic membraneless droplet systems.

A key difference between the membraneless bacterial active droplets developed here and membrane vesicles encapsulating active constituents is that the liquid-liquid interface allows solvent exchange. This essential feature enables prominent hydrodynamic interactions between the droplets and the external fluid environment. Such droplet-environment hydrodynamic interactions may underlie the coupling between the persistent motion of active droplets and the shape deformation due to spontaneous symmetry breaking of internal active stresses. The hydrodynamic interaction may also couple the active stress state between nearby droplets, resulting in the coordinated cellular motion between nearby droplets and giving rise to the anti-phase entrainment of their shape deformations as observed in the experiments. Understanding these hydrodynamic interactions (e.g., in the framework of active phase field models (26, 44, 55) will inform strategies to manipulate the active stress state via this droplet-medium mechanical interaction, thereby allowing controlled patterning and directed transport of active droplets.

**Supplementary Information**, including Supplementary Methods and Figures and Videos, are available in the online version of the paper.

**Data availability.** The data supporting the findings of this study are included within the paper and its Supplementary Materials.

**Code availability.** The custom codes used in this study are available from the corresponding author upon request.

**Acknowledgements.** We thank Daniel Kearns and Harald Putzer for generous gifts of bacterial strains. Y.W. acknowledges support by National Natural Science Foundation of China (NSFC No. T2425007, Y.W.), the Research Grants Council of Hong Kong SAR (RGC Ref. No. 14307822, 14309023, RFS2021-4S04 and CUHK Direct Grants, Y.W.) and by Ministry of Science and Technology Most China (No. 2021YFA0910700, Y.W.), as well as support from New Cornerstone Science Foundation through the Xplorer Prize.

**Author Contributions**: C.L. designed the study, performed experiments and simulations, analyzed and interpreted the data. D.C. and S.L. performed initial experiments. Y.W. conceived the project, designed the study, analyzed and interpreted the data. Y.W. wrote the paper with C.L.'s input.

**Declaration of Interests:** The authors declare no competing financial interests.

**Figures**

**Figure 1**

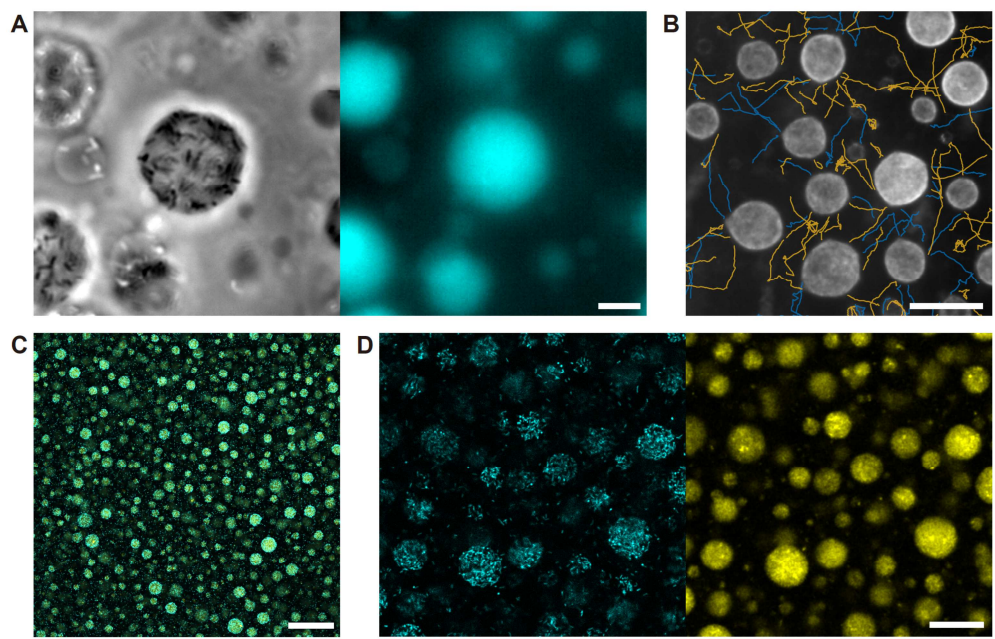

**Fig. 1. Formation of membraneless active droplets encapsulating motile bacteria via liquid-liquid phase separation.** (A) Microscopy image of active droplets encapsulating motile bacterial cells (*B. subtilis* DS91). Left: phase contrast image; right: fluorescent image of the gelatin-rich phase labeled by Oregon Green™ 488 gelatin conjugate. Cell density inside droplets: $6.8\times10^9$ cells/mL; scale bar: 10 µm. (B) Single-cell tracking revealed unbalanced cellular flux entering versus leaving the bacterial active droplets. The cellular flux ratio in this example was measured to be ~1.5:1 (at ~10 min after sample preparation), and the ratio gradually decreased as the cell density inside active droplets increased. Cell trajectories entering (yellow lines) and leaving (blue lines) active droplets were overlaid onto the long-exposure, gray-scaled fluorescence image of GFP-labelled cells (*B. subtilis* DK7921; Methods). The domains with brighter fluorescence (i.e., higher cell density) represent the location of active droplets. Cell density inside droplets: $2.4\times10^9$ cells/mL; scale bar: 50 µm. (C) Enrichment of passive microspheres (0.2 µm diameter, carboxylate modified and fluorescently labelled) in the bacterial active droplets. This panel shows the overlaid fluorescence images from GFP-labelled bacteria (false-colored in cyan) and from microspheres (false-colored in yellow) acquired by confocal microscopy (Methods). Cell density inside droplets: $6.8\times10^9$ cells/mL; scale bar: 100 µm. (D) Zoomed-in view of panel C in separate fluorescence channels (cyan: bacteria; yellow: microsphere) showing the co-localization of bacteria and microspheres. Scale bar: 40 µm.

**Figure 2**

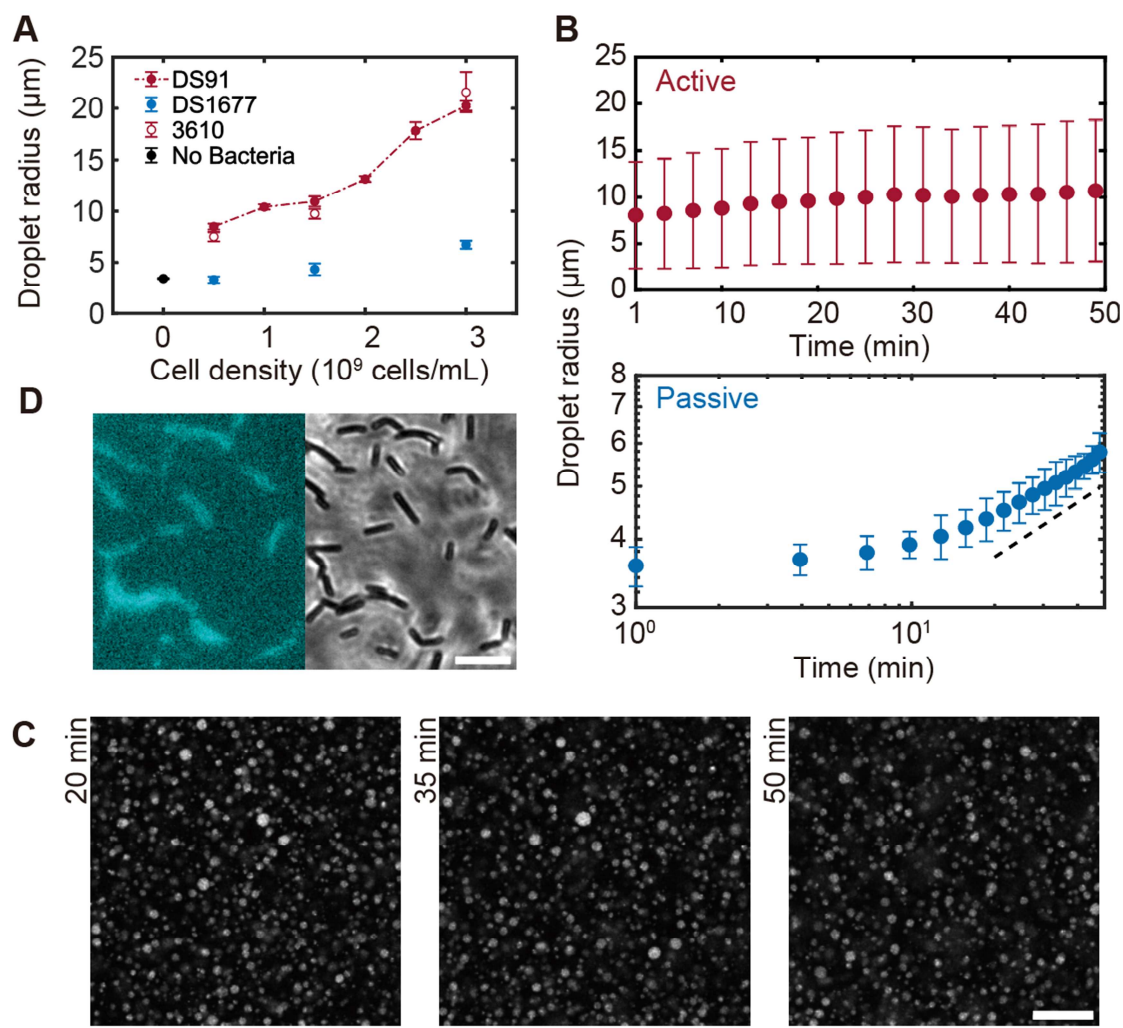

**Fig. 2. Condensation of gelatin-rich droplets depend on bacterial motility.** (A) Dependence of droplet size on overall cell density in the suspension. The mean radius of gelatin-rich droplets $\langle R \rangle$ was calculated as the correlated length of the fluorescence intensity pattern in wide-field fluorescence images of Oregon Green™ 488 gelatin conjugate taken at 10 min after sample preparation (Methods). Bacterial strains *B. subtilis* 3610 and DS91 have wild-type flagellar motility, while *B. subtilis* DS1677 is a non-motile mutant (Methods). Black data point represents droplet radius in the absence of bacteria. Dashed line serves as a guide to the eyes. Error bars represent standard error of the mean (N>4). (B) Temporal evolution of droplet radius in the presence (upper) and absence (lower) of motile bacteria (cell density inside droplets: ~2.0×10$^{10}$ cells/mL; overall cell density in the suspension: 1.5×10$^9$ cells/mL). The mean radius of gelatin-rich droplets was determined by confocal fluorescence microscopy images of 0.2-μm fluorescent microspheres (Methods). Gray dashed line in the lower figure indicates a scaling of $t^{1/3}$; note that the size of passive droplets at the initial stage appeared to deviate from $t^{1/3}$ scaling because a substantial portion of droplets at this stage were smaller than ~0.5 μm in diameter (see Fig. S1) and their size could not be accurately determined by the fluorescence of the few encapsulated 0.2-μm microspheres. Axes for the upper panel are linear,

while those for the lower panel are logarithmic.  Error bars represent standard deviation (N>1900).  (C) Confocal microscopy image sequence of 0.2-µm fluorescent microspheres showing the temporal stability of active droplets.  These images were part of the dataset used in the calculation for panel B.  Scale bar: 100 µm.  (D) Gelatin blobs carried by bacterial cells.  Left: Fluorescent image of the gelatin-rich phase (labelled by Oregon Green™ 488 gelatin conjugate); right: phase-contrast image showing the position of motile bacteria in the same field of view as the fluorescence image.  Scale bar: 10 µm.

**Figure 3**

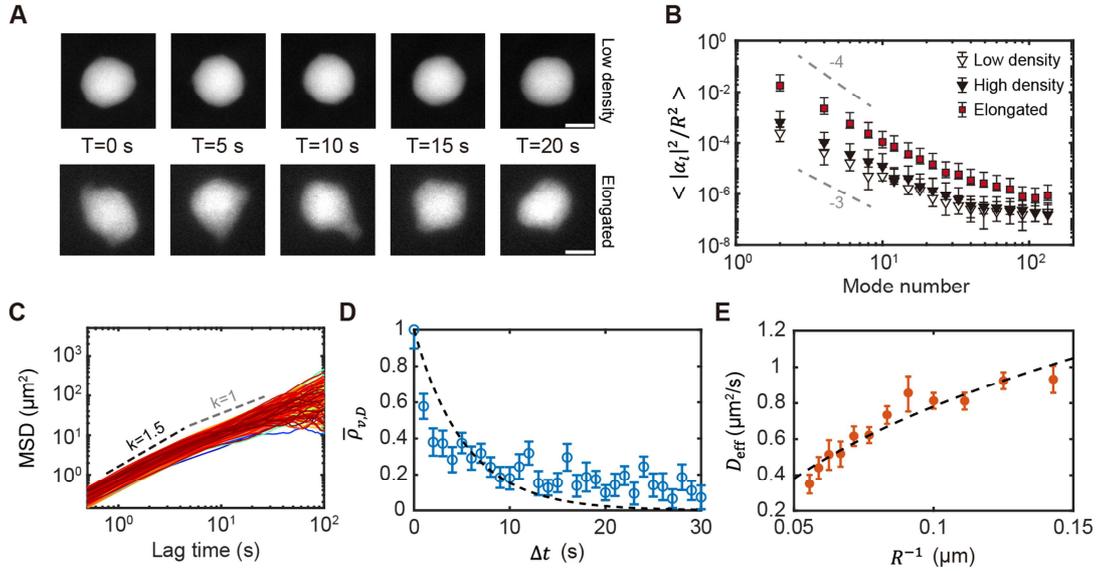

**Fig. 3. Nonequilibrium interface fluctuation and transport of bacterial active droplets.** (A) Image sequences showing the interface fluctuation of bacterial active droplets. The active droplets were labeled by Oregon Green™ 488 gelatin conjugate. Upper row: an active droplet encapsulating *B. subtilis* DS91 (cell length: 4.6±1.2μm; cell density inside droplets: 6.8×10$^9$ cells/mL); lower row: an active droplet encapsulating *B. subtilis* DS3774 (cell length: 10.0±2.8 μm; cell density inside droplets: 1.27×10$^9$ cells/mL) encapsulation. Each row shows five images at T = 0 s, 5 s, 10 s, 15 s, and 20 s. Scale bars: 10 μm. (B) Interface fluctuation spectra of active droplets encapsulating different densities or types of motile bacteria. Open triangles: 6.8×10$^9$ cells/mL *B. subtilis* DS91 (labelled as 'low density'); solid triangles: 2.0×10$^{10}$ cells/mL *B. subtilis* DS91 (labelled as 'high density'); squares: 1.5 × 10$^9$ cells/mL *B. subtilis* DS3774 (labelled as 'elongated'). The rescaled interface fluctuation mode amplitude $\langle |\alpha_l|^2/R^2 \rangle$ was plotted against mode number $l$. Gray dashed lines indicate scaling of $l^{-4}$ (upper) and $l^{-3}$ (lower) as guides to the eyes. Error bars represent standard deviation (N>5 active droplets). (C) Mean squared displacement (MSD) of active droplets encapsulating *B. subtilis* DS91 (cell density inside droplets: 6.8×10$^9$ cells/mL). The black and gray dashed lines indicate slopes of 1.5 and 1, respectively. Colored lines represent the MSD curves of 266 active droplets. (D) Normalized temporal cross-correlation function $\bar{\rho}_{v,D}(\Delta t)$ between instantaneous displacements $v$ and deformation index $D$ (Methods). Normalization was done by dividing the maximum value of the correlation (located at zero delay time). Each data point (gray dots) was obtained by averaging the cross-correlation data from 286 active droplets. Dashed line represents an exponential fit with a correlation time of ~5 s. (E) Effective diffusivity of active droplets $D_{\text{eff}}$ plotted against the inverse of droplets radius $R$. Red data points were obtained from 286 active droplets in total, and the dashed line is the best fit to the data in the form of $D_{\text{eff}} = c_1(R^{-1})^{1/4} + c_2$, where $c_1 = 4.45$ and $c_2 = -1.72$ are fitting parameters. Error bars in panels D,E represent standard error of the mean (N>18).

# Figure 4

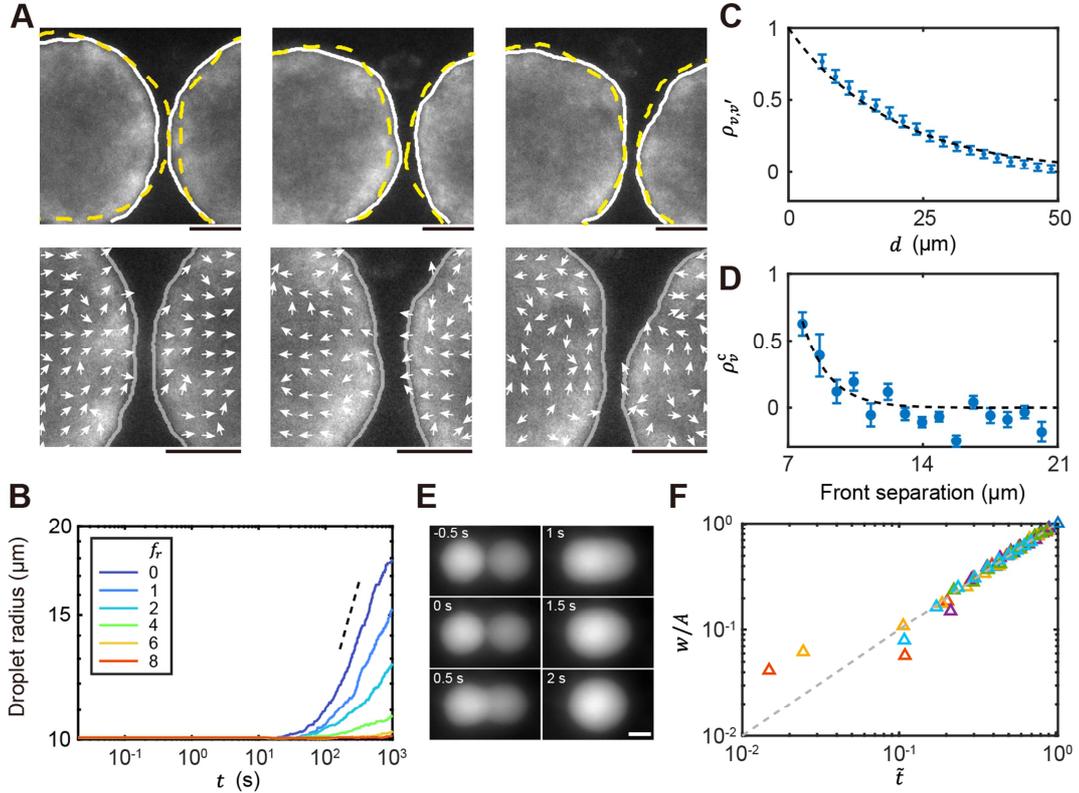

**Fig. 4. Pair-wise interactions of active droplets driven by bacterial activity.** (A) Entrainment of droplet shape deformation and of bacterial collective motion in the droplets. Upper panels are snapshots of two adjacent active droplets (visualized by the fluorescence of GFP-labelled cells *B. subtilis* DK7921 encapsulated in the droplets at a density of ~ 6.8×10$^9$ cells/mL) overlaid with lines indicating the current (white solid line) and subsequent (1.2 s later; yellow dashed line) position of droplet boundaries: left, droplet boundaries moving to the right; middle, droplet boundaries moving to the left; right, droplet boundaries remaining in position. Lower panels: zoomed-in view of the fluorescence images in the corresponding upper panel overlaid with the instantaneous bacterial collective velocity field in the droplets (gray lines show the position of droplet boundaries); the collective velocity field was calculated by particle-image velocimetry technique based on the fluorescence images of bacteria (Methods). Scale bars: 20 μm. (B) Temporal evolution of mean droplet radius in stochastic simulations (see main text and Methods). In the simulation, droplets undergoing diffusive motion would merge upon contact, and an effective repulsion force between the droplets with strength $f_r$ (unit: μm$^2$/s$^2$) tuned the frequency of inter-droplet contacts (Methods). The dashed line indicates a scaling of $t^{1/3}$. (C) Inter-droplet correlation of bacterial collective velocity field (denoted as $\rho_{v,v'}$; see definition in main text). Dashed line represents exponential fit $\sim \exp(-d/L)$ to the experimental data (squares), with the correlation length $L$ being ~ 9.5 ± 2.5 μm (mean ± standard deviation, N=3). (D) Pairwise correlation of droplet centroid

velocity directions (denoted as $\rho_v^c$; Methods) plotted against the inter-droplet front separation. Dashed line represents an exponential fit to the experimental data (circles) with a decay distance (where the correlation equals to $1/e$) of ~8.5 µm. Error bars in panel C, D represent standard error of the mean. (E) Image sequences showing the coalescence of two bacterial active droplets. The active droplets were labeled by Oregon Green™ 488 gelatin conjugate. $T$ = 0 s corresponds to the time of droplet contact. All sub-panels share the same scale bar (20 µm). (F) Log-log plot of neck width versus time during coalescence of active droplets. Color-coded triangles represent experimental data from 6 independent coalescence events. Neck width $w$ was normalized by the average diameter $A$ of the two coalescing active droplets (measured with the cross-sectional wide-field fluorescence images); the time was rescaled and shifted to collapse all data points into the same master curve (denoted as $\tilde{t}$; Methods). Dashed line indicates a scaling of $w = \tilde{t}^{1.0}$. In panels E, F the cell density inside the droplet was ~ $6.8 \times 10^9$ cells/mL.